Dewetting Dynamics of Unstable Lubricant Impregnated Surfaces in Liquid Environment

Abhishek Mund, Arindam Das[*]

School of Mechanical Sciences, Indian Institute of Technology (IIT) Goa, GEC Campus, Farmagudi, Ponda, Goa, 403401, India

*Email: arindam@iitgoa.ac.in

**Abstract:**

This article outlines a thorough stability analysis by means of both theoretical and experimental modeling for Omni-phobic Lubricant Impregnated Surfaces (LIS). The liquid-repellent properties, particularly with regard to water and oil, have gained substantial attention due to their numerous potential applications. One example is Omni-phobic LIS, which shows repellency to oil and water or in a mixed environment of both. The theoretical section of this study focuses on the validation of mathematical models to understand the underlying principles driving the stability of Omni-phobic LIS. Theoretical insights about the interaction of surface texture, chemical composition, impregnated, and ambient liquid properties contribute to a better understanding of the mechanisms that govern stability. A series of experiments were performed to understand better the stability of fabricated Omni-phobic LIS under cyclopentane and water environments, including viscosity and surface texture variations, especially post-spacing variation. The experimental results validate the theoretical predictions and provide valuable statistical information regarding possible model modification. The replaced oil or nucleation sites can be explained through classical heterogeneous nucleation theory (CHNT). The analysis was further validated with experimentally observed nucleation sites. The analysis revealed that these nucleation sites are comparable well with theoretical computations. This confirms the accuracy of the nucleation predictions and supports the underlying theoretical model. According to the Classical Lucas Washburn (CLW) model, the length of a liquid's penetration into a cylinder/square-shaped capillary is expressed as a square root of time. Our findings contribute to designing a stable LIS and then determining the model followed by an unstable one. The proposed Modified Lucas Washburn (MLW) model validated the experimental results. In addition, these experimental data points fit the different capillary imbibition regimes such as inertial, early viscous etc. This contributes to the development of robust and durable solutions for practical applications.

## 1. Introduction:

In nature, liquid wetting and dewetting[1–4] on solids occur naturally and spontaneously. Some natural occurrences, such as a raindrop falling and spreading on leaves, have the fundamental attribute of a liquid drop spreading on a surface[5–7]. Inspired by these environmental processes, scientists/researchers have extensively studied liquid drop spreading on solids to understand the underlying forces better. When a three-phase contact line is present, the equivalent speed at which a liquid can move over a solid surface is extremely limited[8]. Most natural spontaneous imbibition processes involve the interaction of capillary and viscous forces. For quite a while, researchers[9,10] have been investigating how liquid and solid interact. Additionally, it involves more complexities when liquid-liquid and/or liquid-solid interactions are in place, for example, when a liquid is spread on a solid inside another liquid environment. The primary source of fluid flow in a porous medium is typically the capillary pressure resulting from the interfacial tension and sub-mm scale geometry or roughness features of the solid surface. Capillary imbibition using the 2D Lattice-Boltzmann method[11] states that for a low viscous liquid, the dissipation is entirely on the lubricant layer, and the advancing front is in linear growth. As a result, external forces give rise to exponential front growth. It reveals good insight for LIS. The spreading and receding behaviour of tiny water droplets immersed in viscous liquid were experimentally investigated on the textured surface. The researcher[12] developed criteria for the onset of trapped oil under a large droplet and looked into wetting dynamics with a roughness-induced transition from partial to total wetting. At small viscosity ratio, dewetting is slower on low film repellency surfaces due to a high dissipation at the edge of the receding film. This situation is reversed at high viscosity ratios, leading to a slower dewetting on high film-repellency surfaces due to the increased dissipation of the advancing surrounding phase[13]. The dewetting of a dielectrophoresis-induced film into a single equilibrium droplet14.

LIS are textured surfaces that, in certain circumstances, permit a lubricant, such as a liquid, to remain in a thermodynamically stable state within the roughness (liquid or gas). The long-term functional durability of these surfaces, which is necessary for their use in practical applications, depends critically on the retention of this lubricating liquid layer inside the texture of these surfaces over time[15]. LIS is being employed in many practical dynamic scenarios, and the interaction of different fluid properties, including surface tension and viscosity under oil and water environments, poses unique challenges. The flow of one liquid on the textured surface in the presence of another liquid pre-existing inside the texture became more complex and challenging to analyze. Furthermore, the contact line movement takes place with distinct flow directions and time periods. One such example is unstable LIS, where an environmental fluid replaces a lubricant oil present in the texture.

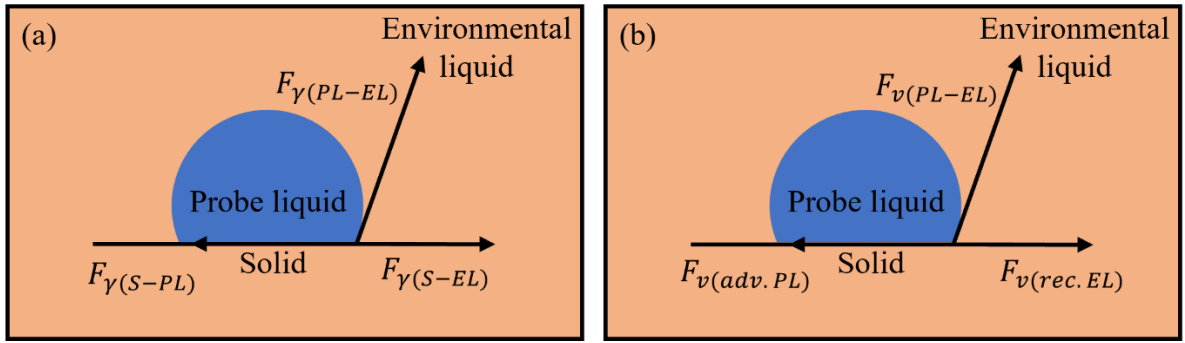

Figure 1. (a) and (b) Schematic of different types of forces for a liquid droplet on a smooth surface while spreading under environmental liquid

$F_{\gamma(S-PL)}$: Surface tension force between solid and probe liquid

$F_{\gamma(S-EL)}$: Surface tension force between solid and environmental liquid

$F_{\gamma(PL-EL)}$: Interfacial tension force between probe and environmental liquid

$F_{v(PL-EL)}$: Interfacial viscous force

$F_{v(adv.\ PL)}$: Viscous resistance force on advancing probe liquid

$F_{v(rec.\ EL)}$: Viscous resistance force on receding environmental liquid

Stable and Unstable LIS has emerged as a promising research area in the field of surface engineering, such as anti-fouling coatings[16], drag reduction systems[17], efficient heat transfer devices[18], and medical devices[19]. The reduction in friction and adhesion facilitates enormous potential applications with stable LIS in an oil/water environment. Some more applications of LIS include asphaltene[20], non-Newtonian fluid[21], drop transport[22], and space applications[23]. Under a dynamic flow scenario, stable LIS under oil and water forms a stable and long-lasting lubricating layer that remains intact. The environmental liquid is unable to replace the impregnated lubricant, resulting in reduced drag and improved efficiency. The impregnated lubricant layer acts as a barrier, preventing direct contact between the solid surface and the surrounding environment. The long-term objective is to conserve energy, which can be attained through drag reduction in a range of industries, including offshore structures, marine transportation, and underwater robotics, by employing surfaces with stable lubricants. In real-world situations, these surfaces have the potential to enhance energy efficiency while additionally improving the equipment's lifespan.

This work broadly discussed the dynamics of imbibition and the displaced lubricant's nucleation rate. Therefore, ongoing research and development focuses on optimizing the design and manufacturing of LIS, paving the way for novel applications in a variety of fields. Theoretical studies play a vital role in providing insights into the fundamental mechanisms governing the stability of oil/water on LIS. The replaced oil or nucleation sites can be explained through classical heterogeneous nucleation theory (CHNT)[24]. It mainly depends on the surface properties, interfacial tension, and the wettability of the surface[25]. This theory is the extension of the classical nucleation theory. It primarily discusses the homogenous nucleation, i.e., occurs uniformly throughout the bulk phase. Applications such as chemical engineering and material science comprehend to understand how surface/interface influences the nucleation process. The Classical Lucas-Washburn (CLW) equation generally describes the liquid penetration length with time[26]. The interactions between the solid surface, lubricant, and the surrounding environment can be explored through mathematical modeling and simulations. It relates the capillary imbibition force through a cylinder or tube with the resistance force. Capillary force is dominant when the adhesive force between solid and liquid is more than the cohesive force between the liquid. There are two basic regimes at play: the inertia and viscous regime[27,28]. These are influenced by several factors, including geometry, surface/interface tension, and fluid properties, namely viscosity. As a result, the CLW equation is often employed to evaluate multiphase flow in porous media. The primary assumption of this equation is incompressible Newtonian fluid follows the Hagen-Poiseuille law. Despite the fact that several methods sought to tackle the challenges in comprehending contact lines (solid, liquid & gas) due to their intricate three-phase phenomena. Wetting liquid viscosity that is often taken into account in the CLW equation. CLW failed to capture the dynamics of the spreading of such environmental fluid in the presence of lubricant inside the texture and deviated from experimental values over an order of magnitude for the current system. Hence, appropriate modifications to the CLW equation are necessary to account for the fluid properties of the impregnated/environmental liquid. The spreading behaviour can be altered because of the impregnated liquid, as it reduces the effective surface tension at the liquid-solid interface. It alters wetting properties, as a result, a Modified Lucas-Washburn (MLW) equation for LIS can be introduced to incorporate these effects.

Following thus, it was found that the length scale had dropped by an order of magnitude. That indicates a greater approximation to the experimental observation. In the proposed modified form, the contact angle of the lubricant oil in the spreading (cyclopentane) phase was considered for the theoretical calculation. The viscosity of the spreading liquid (cyclopentane) used in CLW was replaced with the impregnated liquid viscosity. Following this update in the analytical model, the accuracy significantly improved and became very close to experimental values, i.e., one order of magnitude reduction in the analytical values. This MLW accounts for the presence of the impregnated lubricant, allowing for a more accurate description of the spreading dynamics on LIS surfaces. Furthermore, it was noted that the theoretical model[29] provides strong validation for the distance in relation to the time plot. The fact that the height is independent of viscosity in the initial regime can be explained by the different scaling regimes of capillary rise. This is followed by viscous dominance and the negligibility of inertial effects.

## 2. Theoretical Analysis

### 2.1 Thermodynamic Stability

This theoretical analysis aims to establish a regime map that will specify the conditions for various multiphase interfacial configurations involving three immiscible fluids (including lubricants). Previous studies on LIS stability were limited and solely concerned with lubricant stability under a given probe liquid. Selected lubricants were immiscible with probe liquid and remained inside the texture of LIS under the probe liquid environment. For the theoretical analysis, textured surfaces with square micro posts and a rectangular array were taken into consideration. An optical lithography-based microfabrication technology makes it simple to evaluate and produce this kind of structure. Consider a solid surface that has square posts affixed to it (see Figure 2).

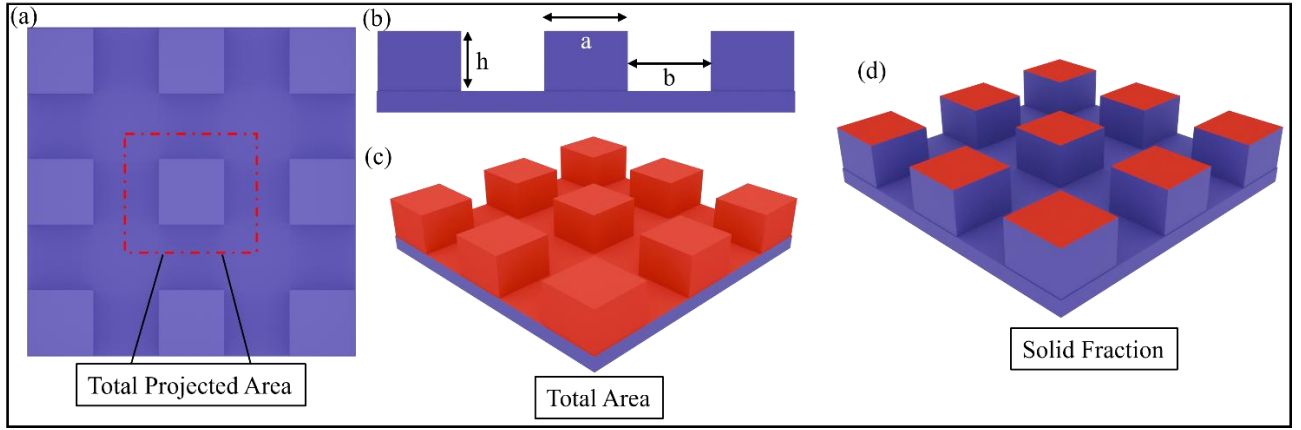

Figure 2. Schematic of representation of square post textured surface (a) Total projected area (b) dimensions (c) Total area (d) Solid fraction

Theoretical analysis is performed considering fundamental geometrical properties such as $r_s$ (the ratio of total area to projected surface area), solid fraction $\varphi$ (the ratio of emerged surface area to projected surface area), width ($a$), edge-to-edge spacing ($b$), height ($h$), and critical contact angle ($\theta_c$). Equations 1, 2, and 3 give the expressions of $r_s$, $\varphi$, and $\theta_c$ for micropost surfaces respectively, as shown earlier.

$$r_s = \left[\frac{(a+b)^2 + 4ah}{(a+b)^2}\right] \qquad (1)$$

$$\varphi = \left[\frac{(a)^2}{(a+b)^2}\right] \qquad (2)$$

$$\cos \theta_c = \left[\frac{1-\varphi}{r_s - \varphi}\right] \qquad (3)$$

The energy minimization principle was used to identify the thermodynamically stable interfacial configuration under different conditions defined by the relationship between surface roughness and surface chemistry parameters. The energy that was minimized here is the overall interfacial surface energy between distinct solid and liquid phases. This energy is the sum of the products of interfacial areas and particular specific interfacial energies of all interfaces between different phases. Geometry and surface chemistry are the sole determinants of these two terms. These specific surface energies (interfacial forces) are a direct articulation of intermolecular forces. These interfacial surface forces balance each other at three-phase contact lines in an equilibrium state.

## 2.2 Spreading Dynamics of lubricant oil

The mechanics of wicking into porous media[30] is fundamentally similar to that of capillary imbibition[31] into a smooth channel. Capillary forces drive the flow, and viscous force resists it, with the balance determining the flow velocity. The complex internal geometry[32] of porous media, size, shape, tortuosity, and wetted part of pores that are difficult to measure exactly pose a significant challenge in comprehending the flow into porous media. When the flow occurs across a porous or rough surface, the structure of the surface may be easily assessed compared to the pores within a bulk. When a liquid comes in contact with a highly wettable surface with sub-millimetric roughness, it impregnates the surface. The stability of impregnated liquid, i.e., retention and/or entrapment, relies on various mechanisms to maintain a continuous liquid film. Capillary forces play a crucial role in retaining the liquid within the textured structures, preventing its depletion. Surface energy minimization, facilitated by the textured morphology and surface chemistry, aids in maintaining the liquid film stability.

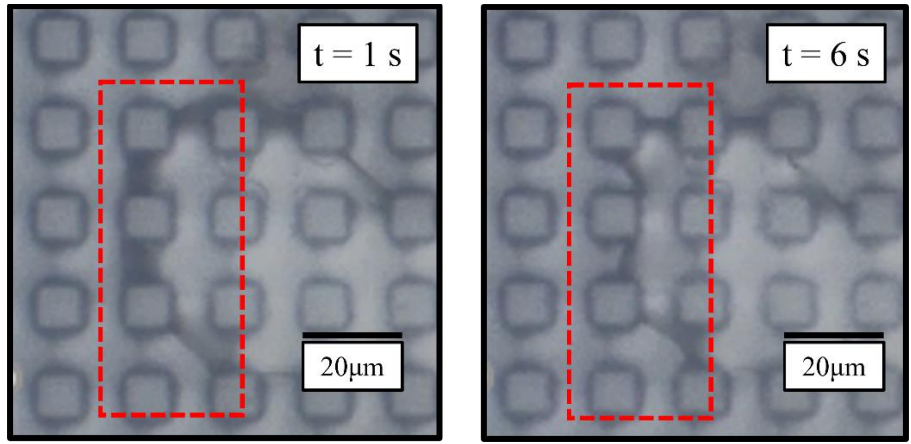

Figure 3. Image sequence of zipping motion. The sample is the square array with 10, 10 and 10 $\mu m$ width, edge-to-edge spacing, and height respectively. Impregnated and environmental oil is krytox 1506 and cyclopentane respectively.

Unstable LIS are characterized by a lack of continuous and persistent liquid film. Instability can be caused by liquid film breakup, induced by external forces or geometric factors. Dynamic wetting transitions[28,33], such as the transition from complete wetting to partial wetting, can further disrupt the stability of the liquid film. The lubricant stability or retention mainly depends on the LIS surface texture and topography. Micro-nano hierarchical structures enhance the capillary forces and promote liquid entrapment, contributing to stability. Surface wettability, characterized by contact angle measurements, influences the stability of LIS. LISs enhance liquid retention and stability by minimizing liquid spreading and promoting complete wetting. Surface energy, determined by surface chemistry and composition, affects the interaction between the lubricant and the textured surface. Interfacial forces, such as van der Waals forces and electrostatic interactions, can influence the stability of the liquid film by promoting liquid spreading or repelling it from the surface. These forces depend on factors like surface chemistry, surface roughness, and the properties of the lubricating liquid.

### 2.2.1 Classical/Modified Lucas Washburn Equation

Since the 1920s, researchers have examined the dynamics of capillary flow in cylinders. Recent studies have been carried out on non-cylindrical uniform geometries[32]. Modern tools use Newtonian or non-Newtonian fluids in some unique applications and have more complex geometries than a homogenous cylindrical shape. Thus, the motivating factor behind this theoretical research becomes apparent. Even if the analysis is quite challenging, studying these fluids' behavior seems intriguing. We attempted to demonstrate the dynamics of triple line movement in this work. One well-known classical phenomenon in the studies of interfacial dynamics is impregnation. In this context, lubricant stability inside the textures of LIS is a critical demand to retain its long-term functional durability.

To have a capillary rise in a tube, the radius needs to be smaller than the capillary length. Similar to other geometry, such as rectangular/square geometry. Capillary/Surface tension force for a rectangular cross-section.

$$F_{st} = 2(b+h)\gamma_{lv}\cos\theta \qquad (4)$$

Where $b$ is the width, $h$ is the height, and $\gamma_{lv}$ is the surface tension between liquid and vapour

Resistance force such as viscous force from Hagen-Poiseuille's equation for a square cross-section

$$F_{viscous} = \frac{\pi^4 \mu}{8\left[1-\frac{2h}{\pi b}\tanh\left(\frac{\pi b}{2h}\right)\right]}\frac{b}{h}s\frac{ds}{dt} \qquad (5)$$

Where $\mu$ is the fluid viscosity, $s$ is the length, $t$ is time.

After balancing and computing equations 4 and 5, resulting in

$$s = \sqrt{\frac{16h\left[1-\frac{2h}{\pi b}\tanh\left(\frac{\pi b}{2h}\right)\right][2(b+h)\gamma_{lv}\cos\theta]t}{\pi^4 \mu b}} \qquad (6)$$

This aforementioned length expression states that the fluid flow in the square-shaped capillaries follows the $t^{0.5}$ scaling law[27].

**2.2.2 Capillary imbibition**

Capillary rise is one of the most well-known and remarkable examples of the capillarity phenomenon, the ability of a liquid to move through small openings without the aid of external forces. The interaction of cohesive forces inside the liquid and adhesive forces between the liquid and the surrounding solid surfaces results in this intriguing behavior. As a result, liquid can ascend against gravity in thin tubes or porous materials, forming a meniscus where the liquid meets the solid boundary. Darcy, for the first time, investigated liquid flow through porous media experimentally. Capillary rise has significant applications and is integral to various natural and engineered processes[34]. In the biological realm[35], it plays a pivotal role in the drinking strategies of insects, birds, and bats. For instance, certain insects exploit capillary action to draw liquid into their feeding structures, while birds and bats may utilize similar principles when feeding on nectar or other liquid sources. This capability is essential for survival, especially in environments where traditional suction mechanisms are inefficient or impractical. Beyond biology, capillary rise is also critical in geophysical settings[36]. It governs the movement of water and other fluids through porous media, such as soil, sand, and rock. Capillary rise is fundamental to understanding fluid transport in small-scale systems from a microfluidics perspective. Microfluidics has attracted a lot of attention because of its numerous applications. It mainly deals with microchannels, which are particularly employed during fluid manipulation. The researcher can analyze fluid behaviour in microscale environment by employing the principles of capillarity. The aforementioned idea is crucial for developing innovative approaches to water resource management and agriculture that ensure effective fluid delivery, appropriate nutrient dispersion, and long-term groundwater replenishment plans.

It was noted that the theoretical model[37] provides strong validation for the distance in relation to the time plot. The graph below illustrates various scaling regimes of capillary rise.

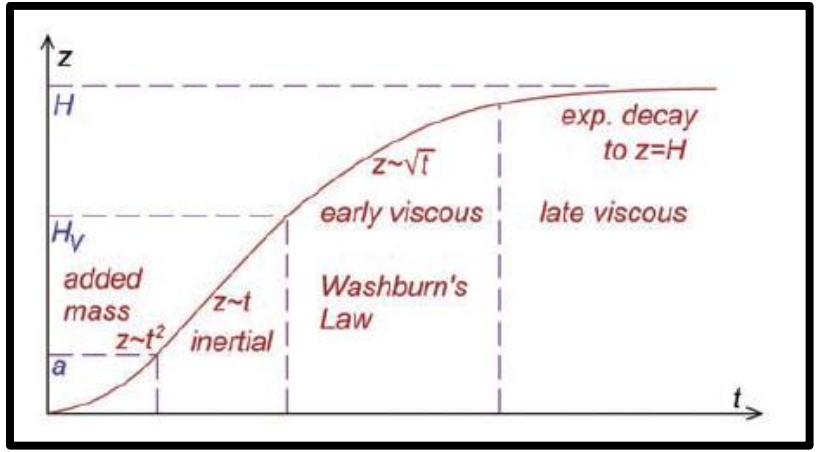

Figure 4. Various scaling regimes of capillary rise[37]

The fact that the height is independent of viscosity in the initial regime can be explained by the different scaling regimes of capillary rise. This is followed by viscous dominance and the negligibility of inertial effects. The independence of height from viscosity in the initial regime arises from the unique scaling laws governing capillary rise, where the interplay of surface tension and gravitational forces primarily determines the dynamics. Here, in this case, the interplay between the surface tension and viscous force. During this phase, the liquid climbs the capillary without significant resistance from viscous forces, as the timescale of their influence is not yet dominant. The magnitude of height is higher in the low viscous case, whereas in the case of high viscous, it is the opposite for the initial time scale. As the process evolves, viscous forces begin to take control, slowing down the rise of the liquid due to the increasing resistance to flow within the capillary walls. This marks the transition to a regime where viscous effects dictate the liquid's movement. Simultaneously, inertial effects, which may have played a role in the very initial stages, diminish in importance, becoming negligible compared to the combined impact of surface tension and viscous resistance. This progression highlights the shift in dominant physical mechanisms as capillary rise advances

## 2.3 Classical Heterogeneous Nucleation theory (CHNT)

Classical Nucleation Theory (CNT), which is primarily concerned with homogeneous nucleation, or nucleation that happens uniformly across the bulk phase. Nucleation is the process by which a small cluster of atoms or molecules (nucleus) creates a new phase inside the parent phase. Classical Heterogeneous Nucleation Theory (CHNT) is utilized to explain how a new phase, such as a solid from a liquid, forms at interfaces or surfaces, including those created by impurities or container walls. Surfaces or interfaces facilitated in the process of heterogeneous nucleation. Activation Energy is the energy barrier that must be overcome for nucleation to occur. The presence of a surface in heterogeneous nucleation lowers this barrier when compared to homogeneous nucleation. In this case, the total interfacial energy per unit area is calculated for state 1, in which the krytox oil is inside the texture. For the same, an appropriate contact angle, along with the specific interfacial energy, was considered. The second state, an environmental liquid (cyclopentane), replaced an impregnated oil (krytox). Factors Influencing CHNT are surface properties, contact angle, and interfacial tensions.

$$E_1 = \gamma_{K1506-FS}(r_s - \emptyset) + \gamma_{FS-CP}(\emptyset) - \gamma_{K1506-air}(1 - \emptyset) \qquad (7)$$

$$E_2 = \gamma_{FS-CP}(r_s) \qquad (8)$$

The driving force can be deduced from the energy calculation provided. The resistance force is influenced by parameters such as geometry and viscosity. In other words, the shape of the object and the fluid's internal friction both play a vital role in how much resistance is encountered. In this particular case, the geometry-induced energy barrier, referred as $(\Delta G)$. The energy barrier can be understood as the amount energy required to overcome the surface interactions, mathematically can be written as the product of surface tension and the area.

$$\Delta G = \delta \Delta A \qquad (9)$$

In this instance, the related area may qualitatively dependent on two specific factors: the receding contact angle and the post spacing. The smallest observed contact angle value before the contact line recedes, considered to be the receding contact angle. The post spacing refers to geometrical parameters i.e. the distance between two consecutive posts (pillars). Both of these factors can significantly affect the area involved in the energy barrier calculation process. In the current analysis, for the sake of simplicity the system is considered as a sphere. This will facilitate a simpler theoretical understanding and calculation process.

$$A = 2\pi r^2 (1 - cos\theta) \qquad (10)$$

By modeling the system as a sphere, we can use spherical geometry to describe the area and thus calculate the energy barrier.

## 3. Result and Discussion

### 3.1 Fabrication of LIS

The designing and fabrication of LIS were performed based on the theoretical computations outlined in the theoretical section[38-39]. When selecting a lubricant, the most important thing to consider is its immiscibility with probe fluids, notably water and cyclopentane. Because liquids that are easily miscible in oils and vice versa are also easily miscible in water, it isn't easy to achieve simultaneous immiscibility with both. The next step is to select a surface chemistry that will modify the surface energy. The common practice to change the surface free energy is to employ silanes on different surfaces such as glass and silicon. Silane reactive sites chemically react with OH-group-containing surfaces, attaching their functional groups to them. Surface energy modifications are implemented using the molecular structures of these functional groups. Fluorosilanization was chosen for further fabrication of LIS surfaces and experimental evaluation. The smooth and critical contact angle correlation determines the thermodynamically stable LIS surface texture. Careful observation reveals that the critical contact angles of FS micropost surface with 5 microns and 20 microns post spacing are just higher than $\theta_{KRY-FS(CP)}$ (72°) and $\theta_{KRY-FS(W)}$ (48°) respectively. On the basis of available fabricated samples, this post spacing can be considered as critical post spacing. To experimentally validate the stability argument, micropost surfaces with 10, 15, 20-micron post spacings were fabricated and used to make LIS. The following steps are to determine the dip coating speed and experimental validation.

**3.1.1 Selection of Dip Coating Speed:** The textured surfaces considered for additional investigations were impregnated with the lubricant (krytox 1506 and GPL 102) using a dip coating method. The dip coated LIS surface possess no excess lubricant layer however the SLIPS surface comprises of thick layer of lubricants present on top of surface texture. The dip coating velocities were meticulously selected in order to prevent an additional lubricant layer. To avoid excess lubricant layer, the capillary number[40] should be less than $10^{-4}$, regardless of surface roughness. The dip coating velocity ($V$) was estimated employing following expression and limiting the capillary number to $10^{-4}$.

$$\text{Capillary number (Ca)} = \frac{\mu V}{\gamma} = 10^{-4} \tag{11}$$

Where $\mu$= liquid viscosity, $\gamma$= surface tension, $V$= withdrawal velocity.

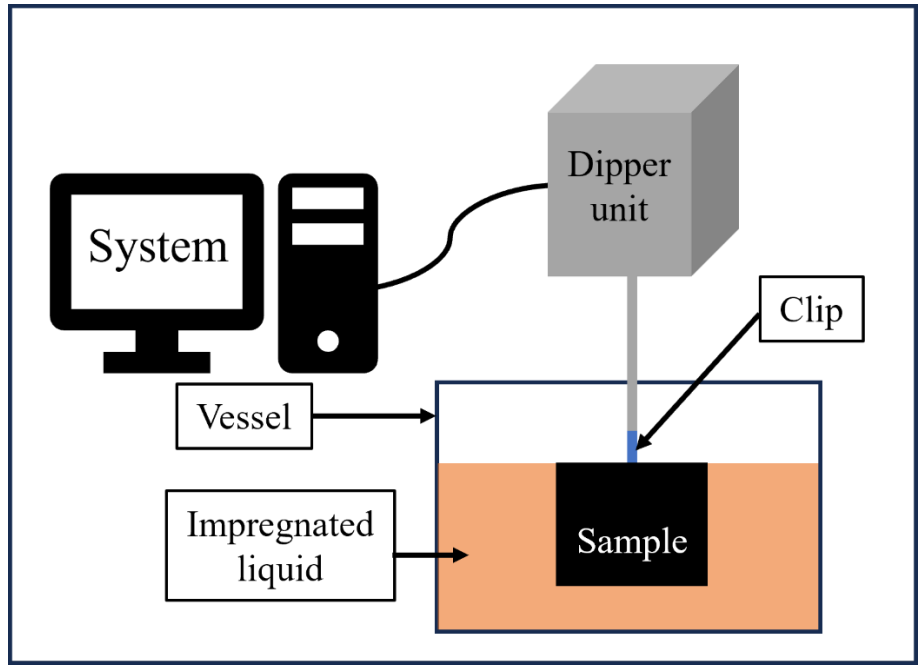

Figure 5. Schematic of the precision dip coating process

Fluid properties such as viscosity, surface tension of krytox 1506 and GPL 102 were tabulated in Table 3.1. The withdrawal/dip coating velocities of krytox 1506 and GPL 102 is $0.9$ $and$ $1.3$ $mm/minute$, respectively.

Table 1. Properties and Withdrawal velocity of lubricant liquid at 20 degrees Celsius[38]

| Lubricant | Liquid-Vapour Surface Tension $(mN/m)$ | Liquid Density $(kg/m^3)$ | Dynamic Viscosity (mPa.s) | Withdrawal velocity $(mm/min)$ | Interfacial Tension $\gamma_{K-CP}$ $(mN/m)$ |
|---|---|---|---|---|---|
| Krytox 1506 | 17 | 1880 | 112.8 | 0.9 | 3.63 |
| GPL 102 | 17 | 1910 | 72.58 | 1.3 | - |

**3.1.2 Experimental validation of Lubricant stability in Micropost LIS:**

The experimental validation was carried out by utilising below experimental setup. Below is the schematic of the experimental setup (see figure 6). The high intensity fiber coupled illuminator with gooseneck Y bundle legs (THORLABS OSL2 Fiber Illuminator) is employed (details in experimental method section).

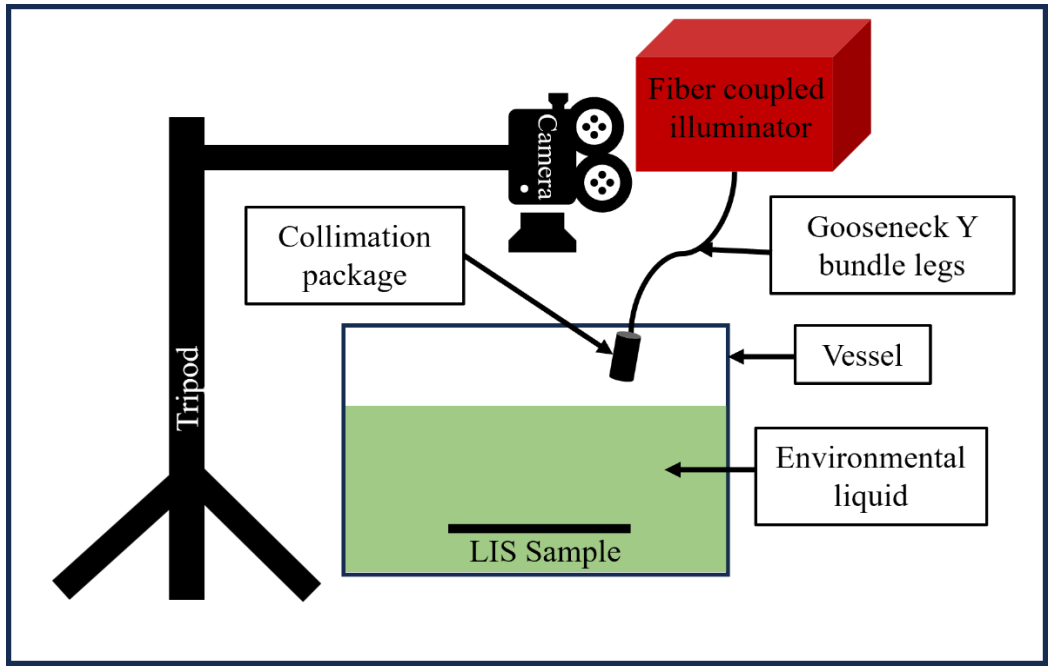

Figure 6. Schematic of the experimental setup

The experimental validation of lubricant stability experiments was investigated on various micro post-LIS in cyclopentane environments. It is evident from the figure below (see figure 7.a, b, and c) that it shows all micron post-spacing LIS samples clearly retained the lubricant at the initial time scale. The samples were kept submerged for long hours; the lubricant instability on 10, 15, and 20-micron post-spacing LIS samples is observed. The environmental fluid cyclopentane replaced lubricant in these samples (Figure 7.d, e, and f). It shows the visible sign of instability. It is evident that the nucleation point[24] is highest in the 10 micron post spacing sample, lowest in the 20 micron post spacing sample, and moderate in the 15 micron post spacing sample for both high and low viscous impregnated systems (see supporting image S1 and S2) essentially at initial time scale.

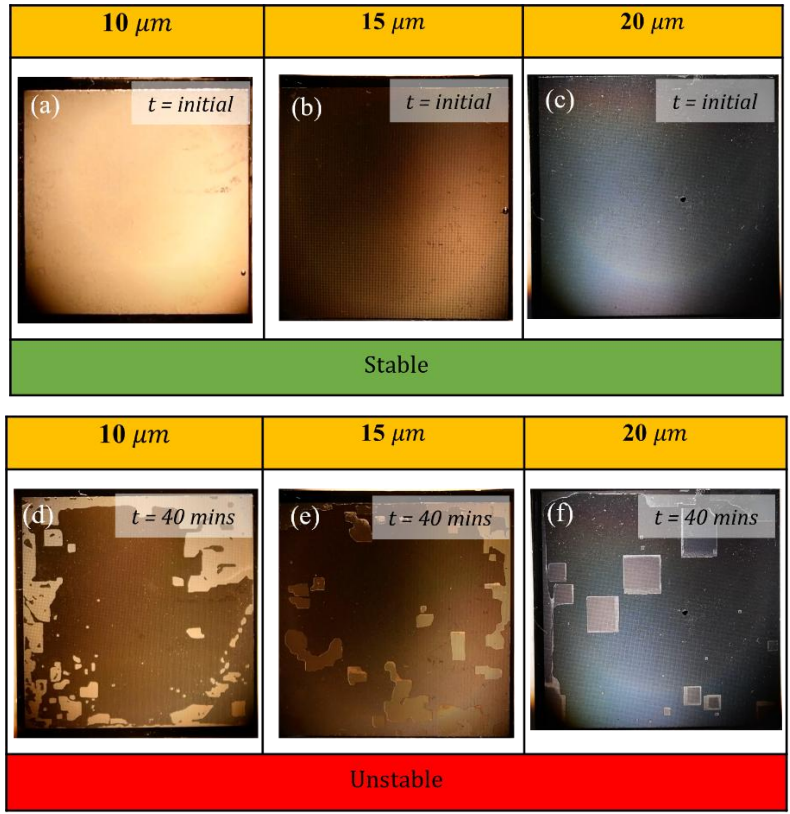

Figure 7. Lubricant (GPL 102) stability images (a) 10μm, (b) 15μm, (c) 20μm post spacing and instability images (d) 10μm, (e) 15μm, (f) 20μm post spacing under CP environment.

The rate at which lubricant is replaced may also be influenced by the fixed sample size. Although cyclopentane is volatile, the environmental liquid level was maintained throughout the experiment.

### 3.2 Spreading Dynamics of lubricant oil

### 3.2.1 Classical Lucas-Washburn Equation

The experimental validation provides evidence that the textured surfaces having post spacings more than critical post spacings are unstable to keep krytox oil inside a given environment fluid. Higher post-spacing surfaces won't be able to hold lubricant in their textures, and lubricant is expected to be replenished by the surrounding fluid. Therefore, when the replacement of lubrication oil is detected, a dynamic scenario that primarily involves the interaction of capillary and viscous force is maintained. The onset time scale of this instability has been found to be inversely proportional to the post spacing. However, in the case of low viscosity, it is directly proportional. The various time scales for lubricant replacement appear to be correlated to the complex dynamics of the three-phase contact line. According to theoretical understanding, the length of a fluid is precisely proportional to the square root of time, according to the CLW equation. The experimental outcomes that were described here follow the same trend.

While validating with experimental data, the velocity that was determined using the CLW equation is in the range of one to two orders of magnitude difference. As wetting fluid viscosity is employed in this expression, the velocity expression will be the same for the low viscous fluid, GPL 102. As cyclopentane is used in our scenario for both experiments, all the velocity values will be the same. When we were validating this, we had considered a MLW expression that could yield results of the same or nearly identical order of magnitude. This observation was slightly expanded by switching the viscosity of the impregnated liquid, krytox, from that of the wetting liquid. The length scale was then found to have dropped by an order of magnitude. That indicates a greater approximation to the experimental observation.

This validation helps us understand the reasons behind the variations in velocity data a slightly clear. The resistance force, which arises in the CLW expression from air to the water/solid surface. Compared to air, cyclopentane's viscosity is quite close, whereas oil's is significantly higher. This indicates that the interaction between these viscosity values produces the cumulative resistance force, this may be the reason why the viscosity of lubricating oil should be taken into consideration. In addition, in this context, the viscous fluid krytox controls the dynamic movement of the triple line. Some more experimental data points are required to understand it more clearly. For better comprehension, more experiments were performed for the initial time scale. A microscopic video with some optical adjustments was carried out to record the dynamics. Because the surface will be immersed in the environment liquid, the working distance that is currently accessible in the microscope is relatively small compared to what is required for this task.

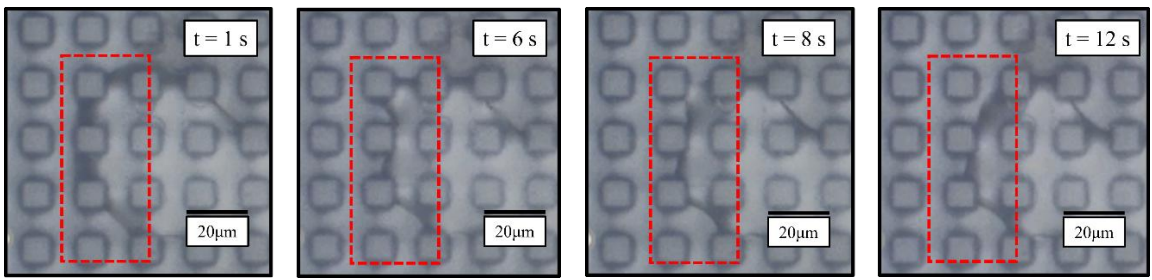

Figure 8. Microscopic images of Krytox 1506 impregnated 10-micron sample under cyclopentane at different time scales.

The facet's motion consists of gradual, quick, and diagonal movements. Somewhere along the front, contact becomes established with a post in the row ahead, and the liquid quickly adheres to its sides. But now the front is deformed, and liquid on each side of the distortion advances and quickly attaches to the post in front. The effect zips down (see supporting video S4) the front in both directions, bringing the entire advancing liquid front forward by one row before the process begins again. In addition, the facet's motion, i.e., side and diagonal, occurs at different time scales.

Table 2. Three-phase contact line motion for different micron post-spacing samples

|  | K 1506 impregnated under CP environment | | |
|---|---|---|---|
| Post-spacing ($\mu m$) | 10 | 15 | 20 |
| Along diagonal ($seconds$) | 5 | 5 | 7 |
| Along parallel to post sites ($seconds$) | 7 | 3 | 3 |

Microscopically evaluating different micron post-spacing samples (see supporting image S3) revealed that facet diagonal and front motion are distinct. Furthermore, it has been noticed that the time scale varies depending on the micron spacing. The time (aforementioned table) was estimated in an initial time scale, which means it is in the inertia regime. This local time is not the average time scale because the velocity at some locations appears to be quick.

### 3.2.2 Modified Lucas-Washburn Equation

The below-mentioned graph illustrates that the experimental data points are in good agreement with the MLW equation. A number of places in the experimental and modified LW equations for the entire micron sample demonstrated improved approximation. Only the lubricating oil's viscosity was considered in this modified LW equation, which fit the experimental results well. Square-shaped geometry is employed in the Hagen Poiseuille equation, which was typically used to balance the capillary force in the classical LW equation. Furthermore, it should be highlighted that the inertial effect predominated during the initial hour (see figure) in the krytox 1506 impregnated sample, causing a sharp increase in velocity during that time. The following section includes another graph that explains the early spreading regime. Viscous dominancy was then observed; however, the velocity decay appeared a little faster than the one with Modified LW.

$$v = \frac{0.5}{t^{0.5}} \sqrt{\frac{16h\left[1-\frac{2h}{\pi b}tanh\left(\frac{\pi b}{2h}\right)\right][2(b+h)\gamma_{lv}cos\theta]}{\pi^4 \mu b}} \qquad (13)$$

Given the lower viscosity, or one order of magnitude less, of samples impregnated with GPL 102 compared to krytox 1506, the velocity will certainly be higher. In the preliminary time scale, inertia force predominates, although, compared to the sample impregnated with krytox 1506, it is more of a capillary and viscous balancing action. The 10-micron impregnated GPL 102 sample validates with the MLW trend, despite the fact that there is essentially one order of magnitude disagreement in the values. The post height could be the cause; the replenished oil could form a paddle, increasing the viscous resistance force. However, 20-micron samples comparatively followed the pattern (see figure 9). The fixed sample size and the level of environmental liquid may play a role determining the instability.

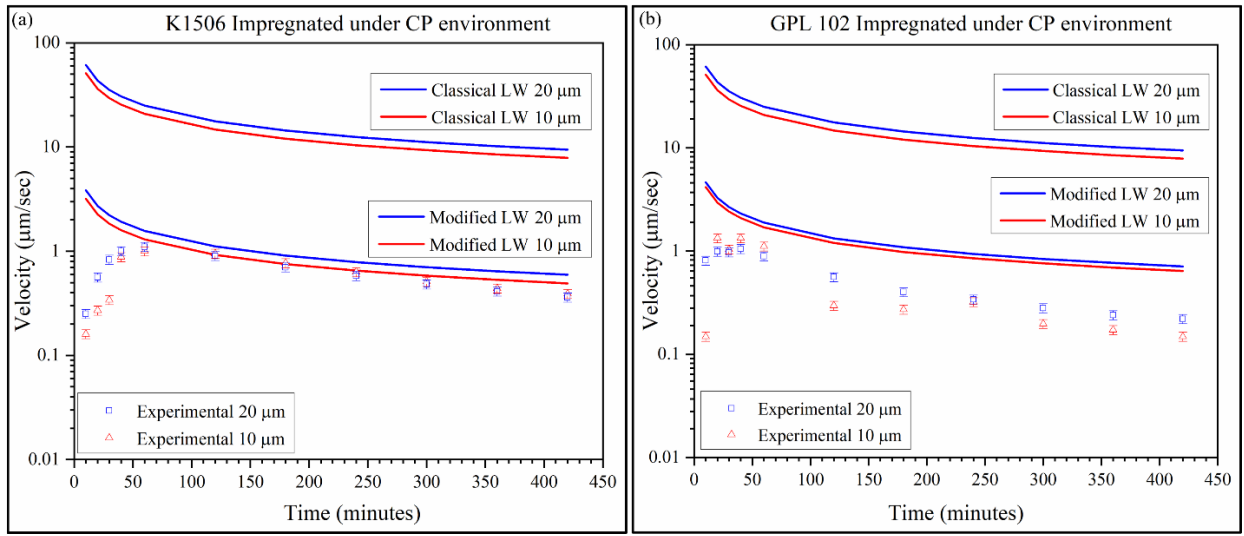

Figure 9. Velocity vs time for (a) krytox 1506, (b) GPL 102 impregnated, various micron sample under cyclopentane environment.

### 3.2.3 Capillary imbibition

Similarly, the finding was made between distance and time for high viscous krytox 1506 and low viscous GPL 102. The distance shown here represents the replacement of lubricating oil under the influence of environmental fluids. The graph below illustrates tanner's law fits, or $s \sim t^n$.

When the graph was validated, it was revealed that the experimental data points follow a capillary imbibition flow regime (see image below). The 15-micron sample (see supporting image S7.1) demonstrates some fluctuation between the added mass region and the inertial regime. Within that location, a 10-micron sample shows higher oil replacement compared to 15-micron samples. The 20-micron sample (see supporting image S7.2) reached faster in the early viscous phase and followed an exponential decay pattern. However, the 20-micron sample traversed more distance than the other samples. The replacement's structure was more like a large square in a 20-micron sample. Each of these samples have been observed attaining the late viscous regime at various time scales. The 10-micron samples were the first to reach the late viscous phase, followed sequentially by the 15-micron and 20-micron samples. This progression suggests a post-spacing dependence on the transition timing. It is experimentally noticed that the 20 micron post spacing sample covers the most distance in the early viscous regime because the post spacing is directly proportional to the distance. In the inertial regime, however, this is not the case.

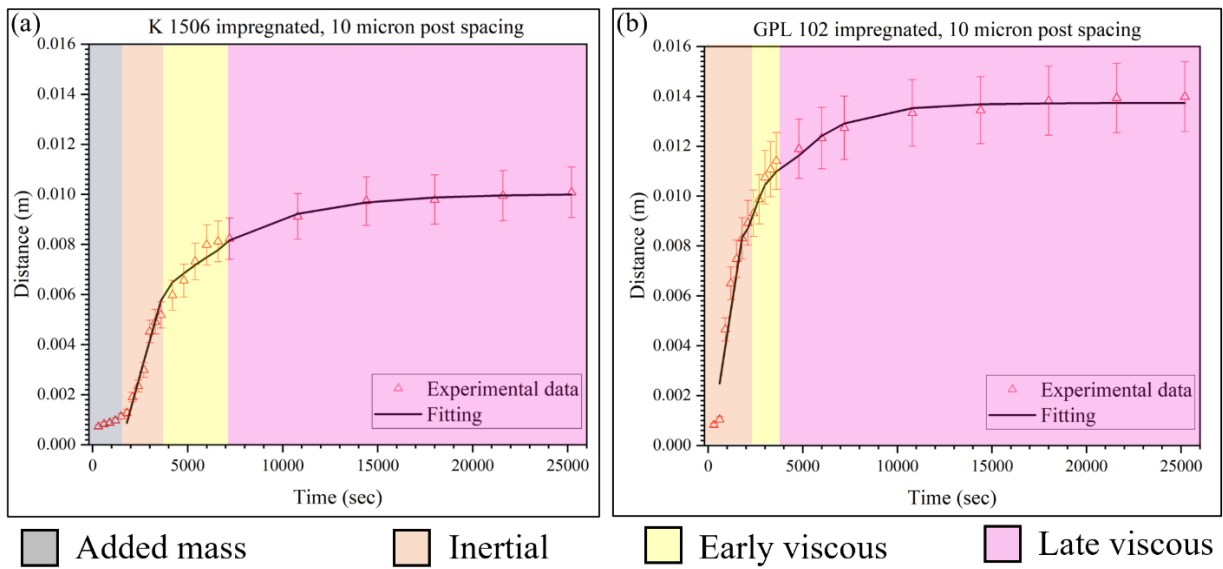

Figure 10. The various scaling regimes of capillary imbibition of distance vs time plot for (a) krytox 1506, (b) GPL 102 impregnated 10-micron sample under cyclopentane environment.

Distance vs time with a lower-viscosity material, namely krytox GPL 102 impregnated was investigated. It demonstrates tanner's law fits, or $s \sim t^n$. Initially, the 10-micron sample had a significantly higher replacement rate than the other two samples. The 15-micron sample shows the least oil replacement throughout compared to the other two. All three samples showed a capillary imbibition flow regime (see the above picture). However, it has been noticed that the distance traveled is inversely related to viscosity. The driving energy for 10 micron post spacing is the same when impregnated with two different viscous oils, however the resistance energy varies. Since inertia is very fast due to low viscosity, it will attempt to attain equilibrium quickly; therefore, the early viscous regime[41] lasts shorter time. The resistance force offered for post spacing, say 10 microns, is relatively low in this situation when compared to higher viscous ones. When low viscosity was compared to high viscosity, the magnitude of distance traveled would gradually increase. In addition, 10 microns indicates the exponential decay faster compared to the other two. The dewetting timescale is controlled by the viscosity ratio and the form of the liquid-liquid interface close to the three-phase contact line. High dissipation near the receding film's edge on low film repellency surfaces enables dewetting to proceed more slowly at low viscosity ratios. In contrast the surrounding phase dissipates more on high film repellency surfaces at high viscosity ratios. Capillary imbibition is primarily resisted by a combination of fluid viscosity, fluid inertia, gravity, and dynamic pressure. According to theoretical analysis, the established initial time scale indicates that viscous effects are insignificant. Initially, fluid inertia mostly resists capillary imbibition. Distance scales with time square in the added mass regime, subsequently with time in the inertial regime. In the early viscous regime, the distance scale with square root of time, as stated in the well-known Lucas Washburn expression. Exponential decay can be observed in the late viscous regime. In addition, it was observed that the experimental data during inertial regime is not much diverged compared to when viscous effect start dominating.

### 3.2.4 Early spreading dynamics

An early time scale study for dewetting between cyclopentane and krytox oil. Three different post-spacing samples were immersed in a cyclopentane bath. Images were captured from recorded video with varying time scales. It is evident that the experimental data points follow Tanner's regime. Power law fittings i.e. $s \sim t^n$, here period of (sec) plotted in following graph.

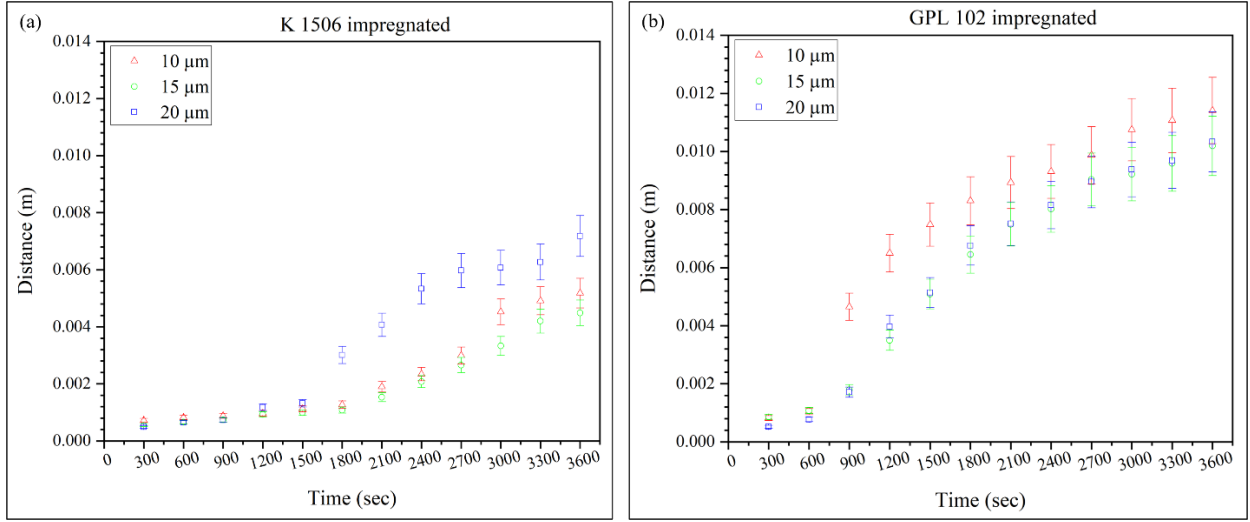

Figure 11. Distance vs time (early time scale) for (a) krytox 1506, (b) GPL 102 impregnated, various micron sample under cyclopentane environment.

For high viscosity krytox 1506, it was found that all post-spacing samples showed nearly identical impregnated oil replacement till close to half an hour. It was observed that at twenty minutes there is a swap in the distance travelled between 10 and 20 micron post spacing sample. The impregnated oil replacement in the experiment for krytox 1506 is validated with different flow regime. Furthermore, the 20-micron post-spacing sample speed up compared to the other two. In contrast, the low viscosity one, krytox GPL 102, shows that impregnated oil replacement is much faster in the early time scale up to twenty minutes. The impregnated oil replacement in the experiment for krytox GPL 102 is validated with different flow regime. After a half-hour time span, impregnated oil replenishment on all samples were very close and was validated.

### 3.3 Classical Heterogeneous Nucleation growth

The theoretical predicted energy was validated using experimental data points. The graph below demonstrates the nucleation point for various micropost spacings with different viscosity impregnated lubricants.

$$\Delta E = E_2 - E_1 \tag{14}$$

The theoretical interfacial energy per unit area value, $\Delta E_{b=10} = 25.099 \ mJ/m^2$, is the highest among all micro-post spacings. The maximum nucleation point was reported for high and low-viscosity impregnated oil at a post spacing of 10 microns. The total roughness and solid fraction are the highest in the case of a 10-micron post-spacing sample. Because of this, the driving energy between the two states is the most prominent. Similarly, the theoretical interfacial energy per unit area value for other post spacing sample, $\Delta E_{b=15} = 19.8149 \ mJ/m^2$ and $\Delta E_{b=20} = 16.8792 \ mJ/m^2$ (see supporting image S5).

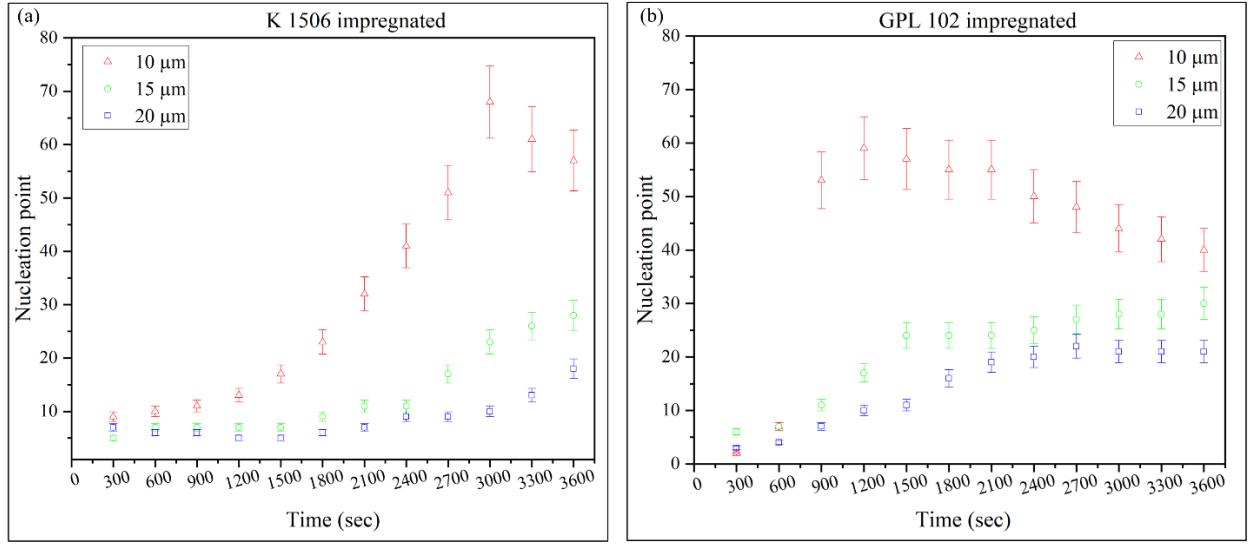

Figure 12. Nucleation point vs time for (a) krytox 1506, (b) GPL 102 impregnated, various micron sample under cyclopentane environment.

For this particular scenario, the theoretical energy barrier that had been mentioned earlier was validated. Qualitatively, it was found that the associated area $A_{b=10}$ is the smallest (see supporting image S6), demonstrating that the 10-micron post-spacing sample had the lowest associated energy barrier when compared to all other samples. The 10-micron post-spacing sample has the highest theoretical driving force and the lowest barriers. The argument is validated by results from experiments showing that the experimental nucleation point is the highest. From a 2/3D view (see supporting image S6), it is observed that the sagging is the least in this sample compared to other post-spacing samples. It means the nucleation energy in the 10-micron post-spacing will be the highest. Similarly, other post-spacing samples, such as 15 and 20 microns, followed the trend. There are relatively more nucleation sites for low viscosity GPL 102. One factor that may contribute to a high nucleation rate is the pace at which distinct nucleation sites merge. This distinct nucleation sites are merging at different time scale. Furthermore, the equilibrium state is reached much more quickly.

## 4. Conclusions and Future scope of work

In conclusion, the investigation of the dewetting dynamic on the LIS surface was experimentally validated using the theoretical model. The theoretical model, i.e., modified Lucas Washburn, fits the experimental data points at an early spreading time scale. It was valid for high and low viscosity impregnated oils in cyclopentane environments. Viscosity and dynamic contact angle significantly impact the stability of LIS. The theoretical calculations for different capillary imbibition regimes are validated by the experimental data points. For both high and low-viscosity-impregnated oil, every sample with a different post spacing fits into a different capillary imbibition regime. The experimental data is in good accord with the inertial, early, and late viscous scaling regimes. An added mass regime is shown in the case of a high-viscosity-impregnated sample, however, not in the case of a low-viscosity one. It is simply because of its low viscosity that it moves quickly. It was observed that the duration of the early viscous regime is shortened in low viscous impregnated materials as the post spacing increases. The reason why early viscous regimes are being squeezed out is because they are achieving equilibrium faster in low viscous impregnated oil. In addition low viscosity impregnated oil has a high inertia. Furthermore, it is found that the early viscous duration is comparatively short in low viscous impregnated oil when comparing high and low viscous impregnated oils with every post spacing. When a highly viscous impregnated sample is used, the 20-micron post-spacing sample travels the farthest. There is a temporal variation, and a magnified view (under a microscope) of the triple line movement demonstrates that it is similar to zipping or unzipping motion. It would be interesting to see the same phenomenon with a different geometry, such as a hexagonal form or another analogous shape. However, additional investigation is required to highlight some more interesting points. It is evident that the stability analysis necessitates consideration of the surface properties. Nucleation points calculated with respect to time were further verified using experimental data points, and it was discovered that these nucleation sites validated with the theoretical computation.

## 5. Experimental methods

**5.1 Sample Cutting:** The test samples were cut from 6-inch diameter silicon wafers that were smooth or textured. A 1064 nm Nd:YAG solid-state laser (Electrox) was used to cut wafers into 20 mm by 20 mm squares. Appropriate designs have been developed employing the laser system's software to cut wafers into the optimum number of samples with the aforementioned dimensions.

**5.2 Surface Chemical Functionalization:** Solid surfaces, both textured and non-textured, were chemically modified using a well-established silane process. A low surface energy silane, FS (*1H,1H,2H,2H-Perfluorooctyl-trichlorosilane*, Sigma Aldrich), was employed for this purpose. Initially, square-cut silicon wafers were plasma cleaned for ten minutes in an RF plasma cleaner (Harrick Plasma, Model PDC-002-HP) in an oxygen-rich environment at 200 mTorr pressure. Following this, a vacuum-based chemical vapor deposition of fluorosilane was conducted. The samples were cleaned in the plasma cleaner for 10 minutes and then placed inside a desiccator containing 5 μL of fluorosilane at 300 mbar for 4 hours. After the chemical reaction was complete, the samples were sonicated separately in acetone (ACS grade, 99%-Sigma Aldrich) and isopropyl alcohol (ACS grade, 99%-Sigma Aldrich) for two to three minutes each, followed by thorough rinsing with DI water. These cleaning methods were used to remove physically deposited silanes from the substrate surfaces.

**5.3 Surface Texturing:** Micro-posts with surfaces on 6-inch silicon wafers were fabricated using standard optical lithography (n-type 1 0 0 planes, 650μm thick). Square micro-posts sized 10μm x 10μm x 10μm were fabricated with inter-post spacings ranging from 5μm to 100μm. The silicon wafer was coated with Shipley S1818 photoresist, which was eventually exposed to ultraviolet light of 405 nm wavelength through a chrome mask (Advance Reproductions Corporations). The exposed wafers were prepared employing a 1:1 volume ratio of Microdev solution (Dow Chemicals) and De-Ionised water. The inductively coupled plasma etching system (ICPES; Surface Technology Systems) was employed for etching process (etched height of 10 μm) for the wafers with a developed photoresist layer. Surface profilometry with a non-contact optical profiler (CCI HD Optical Profiler, Taylor Hobson) validated etched height. Furthermore, the etched samples were cleaned by means of piranha solution (1:3 volume ratio of hydrogen peroxide and sulfuric acid) to get rid of all remaining photoresists.

**5.4 Microscopic Visualization:** An optical microscope (Hirox-RH 2000, 50X objective) was employed to capture microscopic videos. Furthermore, a field emission scanning electron microscope (FESEM) (Sigma 300-Carl-Zeiss) was used to analyze the textured surfaces. A typical image was obtained at a magnification of approximately 1500x and a voltage of approximately 5kV.

**5.5 Contact Angles Measurement:** A contact angle goniometer (Rame-Hart model 500-U1) was employed for determining various contact angles, including equilibrium, receding, and advancing, on variety of surfaces in air, water, and cyclopentane environments. Test samples were submerged in a liquid-filled transparent quartz cell to measure contact angles. Probe liquids were limited to capillary lengths of approximately 4 µL. The releasing and withdrawal rate of 0.2 µL/s was employed to evaluate the advancing and receding angles. At least ten measurements were obtained from distinct locations on the sample to calculate the average and standard deviation of measured quantities.

**5.6 Spreading experiment:** The semi-dynamic investigation was carried out within a transparent quartz cell. This typical quartz cell dimensions $5.5x5x5\ cm$. The quartz cell was deeply cleaned with acetone and isopropyl alcohol solvents before being filled with cyclopentane, sometimes known as environmental oil. A particular micro post spacings LIS surfaces is then immersed in it. A DSLR camera (Nikon D850) with a 40 mm micro lens was used to capture the videos. The high intensity fiber coupled illuminator with gooseneck Y bundle legs (THORLABS OSL2 Fibre Illuminator) is employed. Image analysis of these videos was performed using Fiji (Image J) to measure the distance.

**5.7 Interfacial tension measurement:** The sigma (Biolin Scientific: OneAttention Sigma 700/701 Force Tensiometer) main unit was turned on 24 hours prior performing the measurement. The probe Du Noüy ring properly cleaned with the solvents (acetone, Isopropanol), in addition a standard vessel was thoroughly cleaned. The corresponding program was employed for basic setup, such as specifying the heavy and light phase materials such as krytox oil (heavy phase) was poured in the vessel followed by cyclopentane (light phase). The control parameters must be configured properly. To compute the data, a graph associated with a table will be generated on the system.

**5.8 Dip coating process:** To produce a number of LIS surfaces, the dip coating procedure was employed to impregnate lubricant into all textured FS functionalized surfaces. The dip coating procedure restricts the formation of a thick lubricating layer regardless of surface roughness characteristics. This method was carried out using a precision dip coater (Biolin Scientific: KSV NIMA multi vessel). A conventional vessel was thoroughly cleaned using solvents such as acetone and isopropanol, and subsequently filled with lubricant oil. A functionalized rough surface mount on the cantilever allows it to be immersed in the lubricant at a specified Y speed (mm/min). Once the sample is fully submerged, provide the corresponding withdrawal velocity and carry out experiment.

## ASSOCIATED CONTENT

**Supporting Information**.

The following files are available free of charge

## AUTHOR INFORMATION

**Corresponding Author**

* Email: arindam@iitgoa.ac.in

**ORCID ID**

Abhishek Mund: 0000-0003-2924-1804

Arindam Das: 0000-0002-8163-0666

## AUTHOR CONTRIBUTION

Abhishek Mund: Designed and performed experiments, Theoretical calculations, Data analysis, Writing original manuscript. Arindam Das: Developed the experimental design and concept, Review and editing the manuscript, Supervision. The final document of the article has received the unanimous approval of all writers.

**Funding Sources**

This research was partially funded by the startup grant (Grant No. 2020/SG/AD/030) by the Indian Institute of Technology Goa. Abhishek Mund acknowledges for the financial assistance received under Ministry of Education (MoE), Government of India.

**Notes**

The authors declare no competing financial interest.

## ACKNOWLEDGMENT

The authors acknowledge CoE-PCI (Centre of Excellence- Particulates, Colloids and Interface) of IIT Goa for providing the instrumentation and workspace for the research study. The authors acknowledge CNS - Center for Nanoscale Systems of Harvard University for providing cleanroom facilities for lithographic sample preparation.

**Supporting Image S1. Captured instability at a larger time scale for krytox 1506 impregnated different micron post-spacing surfaces under CP environment.**

These images provide essential insights into the behavior of oil-infused surfaces under oil environments, allowing an overall perspective of the instability between two immiscible fluids. The textured surfaces featuring various micron post spacings were impregnated with krytox 1506 (high viscous) oil, which was subsequently kept inside the cyclopentane. The macroscopic photographs were captured with a DSLR and an optical light source. The images below (figure S1.1) correspond to a 10-micron post-spacing sample. It illustrates stability at the initial time scale, followed by instability at subsequent time intervals. The bright patches (arbitrary shapes) on these LIS surfaces indicate instability. The lubricating oil was gradually replaced by the environmental fluid cyclopentane in certain areas. The interplay of capillary and viscous force mostly determines the dynamics of this process.

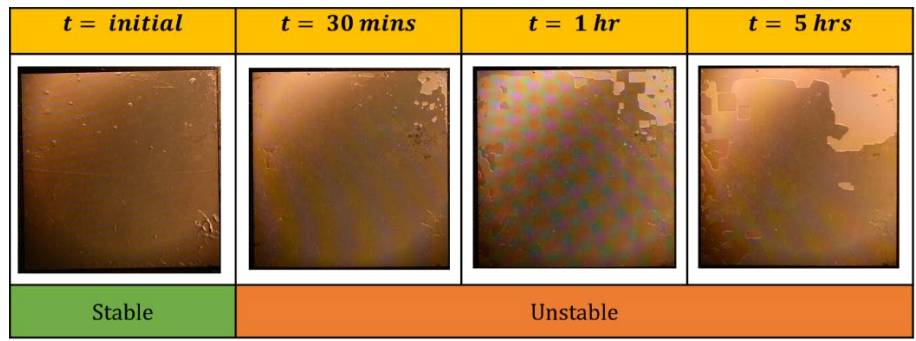

Figure S1.1: Shows macroscopic image of K 1506 impregnated 10-micron post-spacings LIS

The macroscopic images shown below (figure S1.2) illustrate krytox 1506 impregnated for a 15-micron post-spacing sample in cyclopentane oil. Here, the dark patches portions appear to be more rectangular in shape. Impregnated oil replacement was minimal until an hour, after which it increased substantially.

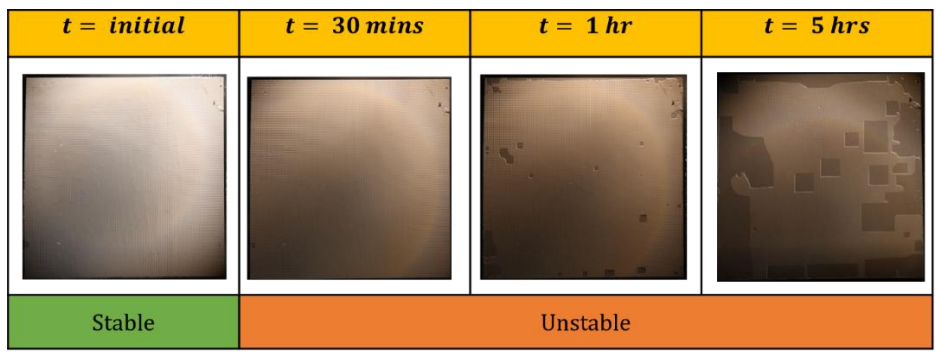

Figure S1.2: Shows macroscopic image of K 1506 impregnated 15-micron post-spacings LIS

The 20-micron post-spacing sample from the initial time scale demonstrated oil replacement. The macroscopic photos below (figure S1.3) show krytox 1506 impregnated for a 20-micron post-spacing sample in cyclopentane oil. A large rectangular shape formed when the impregnated oil was being replaced.

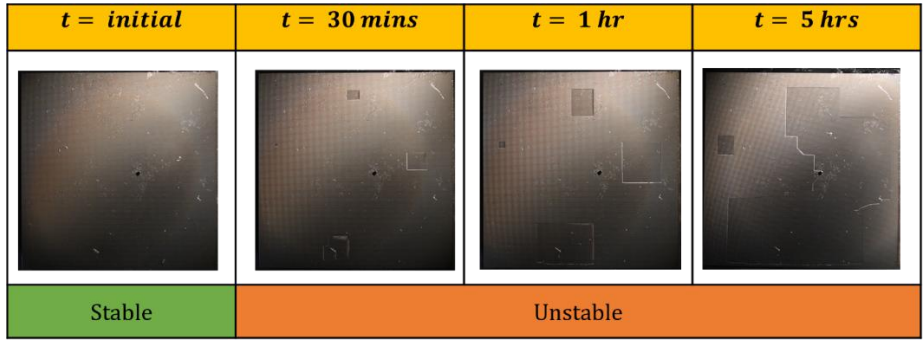

Figure S1.3: Shows macroscopic image of K 1506 impregnated 20-micron post-spacings LIS

**Supporting Image S2. Captured instability at an early time scale for krytox GPL 102 impregnated different micron post-spacing surfaces under CP environment.**

The textured surfaces with varying micron post spacings were impregnated with krytox GPL 102 (low viscosity) oil, which was then kept inside the cyclopentane. The images below (figure S2.1) indicate a 10-micron post-spacing sample. The photos were captured at an early time scale. It is apparent that impregnated oil replenishment has been observed since the initial time scale. Small arbitrary shapes are generated first, and they eventually lead to considerable rectangular shapes. These bright spots indicate areas where impregnated oil was replaced. It is the fastest of all micron post-spacing samples impregnated with a low-viscosity oil.

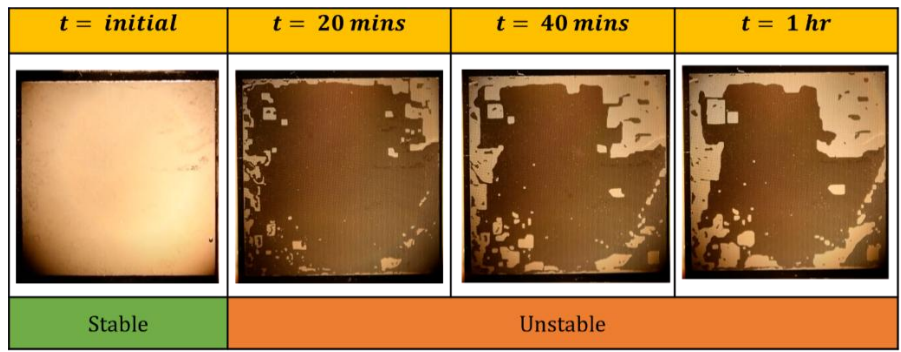

Figure S2.1: Shows macroscopic image of GPL 102 impregnated 10-micron post-spacings LIS

In the 15 micron post-spacing sample, the nucleation spots (impregnated oil replacement) appear to be fewer than 10 microns. The macroscopic images shown below (figure S2.2) illustrate krytox GPL 102 impregnated for a 15-micron post-spacing sample in cyclopentane oil. The shape of the oil replacement is rectangular and small in size.

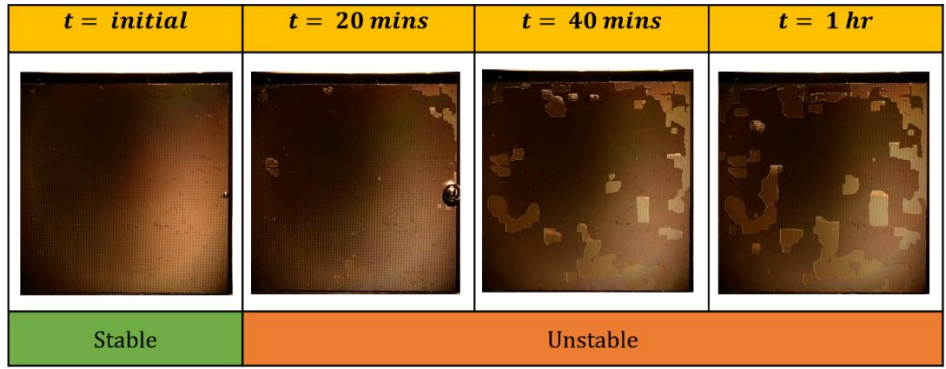

Figure S2.2: Shows macroscopic image of GPL 102 impregnated 15-micron post-spacings LIS

A large rectangular shape formed when the impregnated oil was being replaced. The 20-micron post-spacing sample from the initial time scale displayed oil replenishment, fewer nucleation points, and almost all rectangular shapes. Figure S2.3 shows macroscopic photographs of krytox GPL 102 impregnated with cyclopentane oil for a 20-micron post-spacing sample.

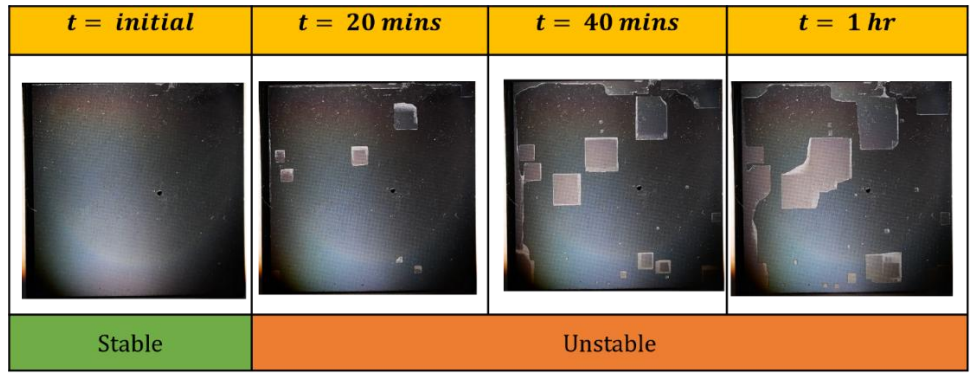

Figure S2.3: Shows macroscopic image of GPL 102 impregnated 20-micron post-spacings LIS

**Supporting Image S3. Captured microscopic images at different time scale for krytox 1506 impregnated different micron post-spacing surfaces under CP environment.**

The microscopic image of different micron samples demonstrates an intriguing world at the micro-nanoscale, offering intricate details previously undetectable to the naked eye. Each pixel in the image depicts just a fraction of the sample's surface, revealing the topography, texture, and composition with remarkable clarity. The microscopic images were obtained with an optical microscope and a 50X objective lens. The textured surfaces with varying micron post spacings were impregnated with krytox 1506 (high viscous) oil, which was afterward kept inside the cyclopentane. The images below (figure S3.1) correspond to a 10-micron post-spacing sample.

The facet movement has been found to be a combination of diagonal and front-to-front, as well as quick and slow at times. It has been demonstrated how facet movement occurs at various time scales.

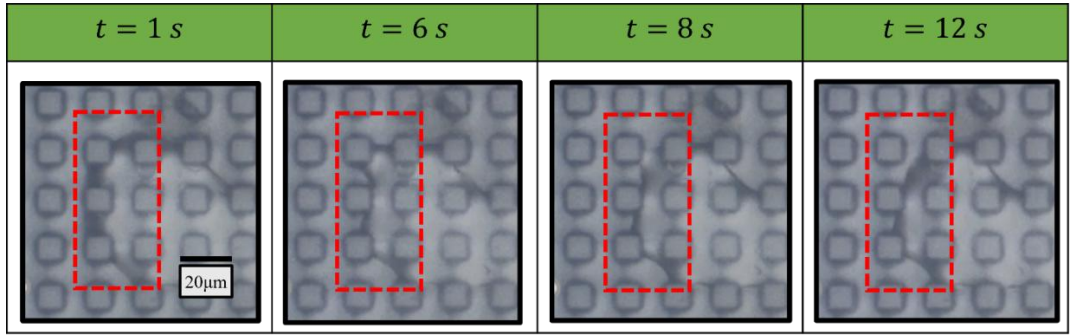

Figure S3.1: Shows microscopic image of K 1506 impregnated 10-micron post-spacings

The illustrations below (figure S3.2) depict a 15-micron post-spacing sample. The facet movement can be observed both diagonally and frontally. As the microscope moves, it shows a remarkable scope and complexity. The dark portion of the photograph depicts the environmental fluid, while the light part represents the impregnated fluid. Instability can be observed at various time scales.

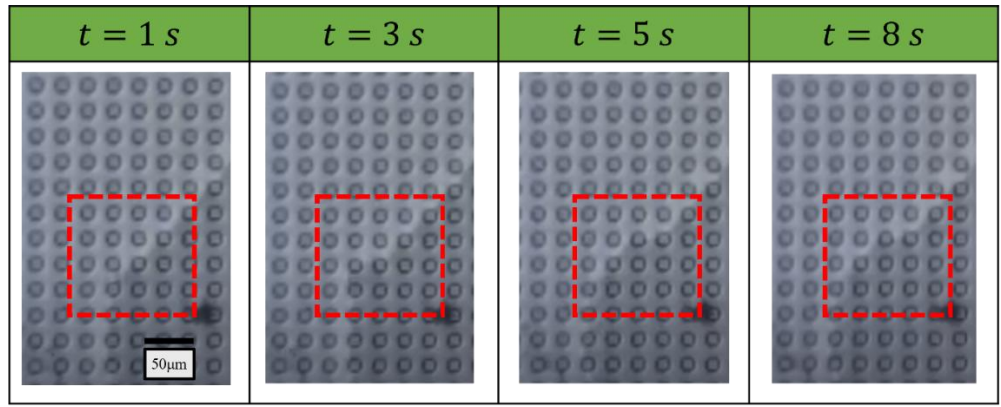

Figure S3.2: Shows microscopic image of K 1506 impregnated 15-micron post-spacings

Microscopic photographs of a micron sample under an oil environment show an evolving environment characterized by interfacial tension and viscous resistance forces. The illustrations below (figure S3.3) depict a 20-micron post-spacing sample. The de-wetting time scale variation can be observed.

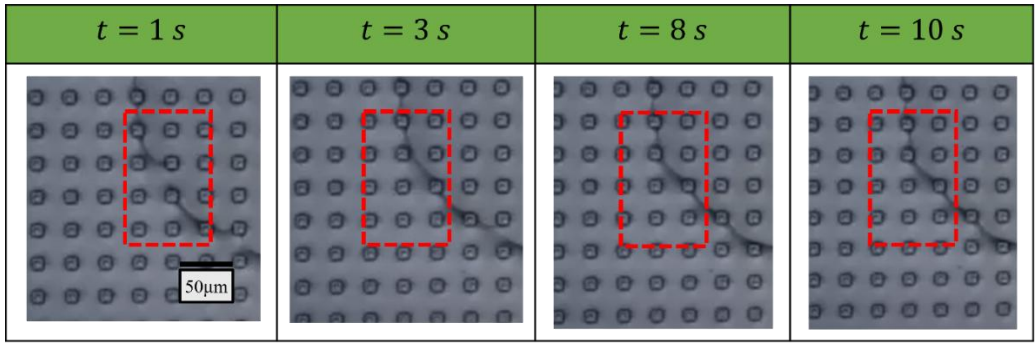

Figure S3.3: Shows microscopic image of K 1506 impregnated 20-micron post-spacings

**Supporting Video S4. Facet movement in diagonal and front-to-front**

The facet movement in the diagonal and front-to-front direction for a 10-micron post-spacing LIS surface under a cyclopentane environment.

**Supporting Image S5. The theoretical interfacial energy calculation with two different states inside the cyclopentane environment.**

The functionalized micro-textured surface impregnated with krytox oil. When the surface is submerged inside a cyclopentane environment, it is expected that on unstable conditions, cyclopentane replace the krytox oil. Here theoretical calculation has been carried out to determine the associated energy (heterogeneous nucleation). Appropriate contact angles and specific interfacial energy parameters were used in energy calculations.

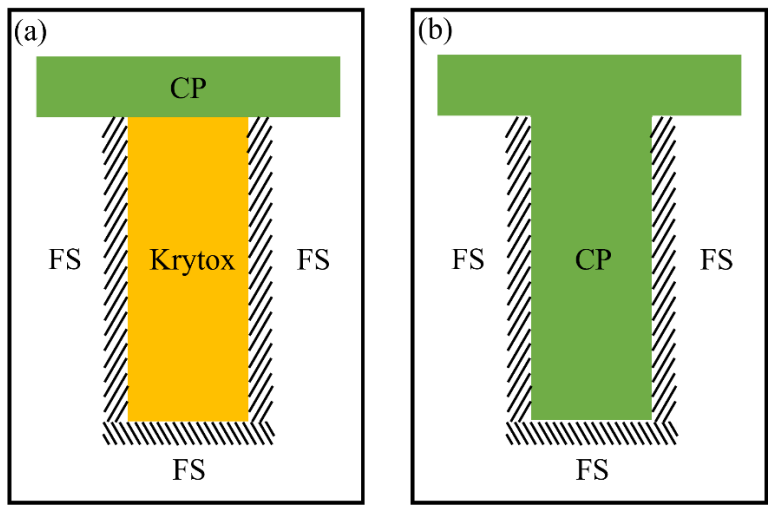

Figure S5. Schematics of a FS functionalized textured surface filled with (a) krytox oil, (b) cyclopentane

The interfacial energy per unit area associated with S5.(a)
$$E_1 = \gamma_{K1506-FS}(r_s - \emptyset) + \gamma_{FS-CP}(\emptyset) - \gamma_{K1506-air}(1 - \emptyset)$$
The interfacial energy per unit area associated with S5.(b)
$$E_2 = \gamma_{FS-CP}(r_s)$$

The receding surface energy[1] $\gamma^{total}(FS) = 26 \ mJ/m^2$ on Trichloro (1H, 1H, 2H, 2H perfluorooctyl) silane surface. The krytox 1506 surface tension[2] $\gamma_{K1506} = 17 \ mN/m$. The measured contact angle of krytox on FS surface $\theta_{K1506-FS-air} = 33.4°$. The interfacial energy between krytox and cyclopentane $\gamma_{K1506-CP} = 3.63 \ mN/m$.

$$\gamma_{K1506-FS} = \gamma_{FS-air} - \gamma_{K1506-air}\cos\theta = 11.8076 \ mN/m$$
$$\gamma_{FS-CP} = \gamma_{K1506-FS} - \gamma_{K1506-CP}\cos\theta = 12.9293 \ mN/m$$

The theoretical interfacial energy per unit area for state $E_1$ with post spacing 10 $\mu m$ calculated to be 0.7596 $mJ/m^2$, similarly for state $E_2$ 25.8586 $mJ/m^2$. The difference between these two points calculated as $\Delta E_{b=10 \ \mu m} = 25.099 \ mJ/m^2$

$E_{1 \, b=10 \ \mu m} = 0.7596 \ mJ/m^2 \quad E_{2 \, b=10 \ \mu m} = 25.8586 \ mJ/m^2 \quad \Delta E_{b=10 \ \mu m} = 25.099 \ mJ/m^2$

$E_{1 \, b=15 \ \mu m} = 1.3891 \ mJ/m^2 \quad E_{2 \, b=15 \ \mu m} = 21.2040 \ mJ/m^2 \quad \Delta E_{b=15 \ \mu m} = 19.8149 \ mJ/m^2$

$E_{1 \, b=20 \ \mu m} = 1.7389 \ mJ/m^2 \quad E_{2 \, b=20 \ \mu m} = 18.6181 \ mJ/m^2 \quad \Delta E_{b=20 \ \mu m} = 16.8792 \ mJ/m^2$

**Supporting Image S6. The energy barrier calculation**

The schematic illustration below shows how a specific post-spacing sample in this case, 10 and 20 microns post spacings is shown. The thin layer (a profile) on top shows how environemtal liquid replaces the impregnated lubricant. In qualitative terms, the related area was calculated using the assumption that the profile is spherical.

For 10-micron post spacings sample the area and the associated energy can be expressed as
$$A_{b=10 \ \mu m} = 2\pi r^2_{b=10 \ \mu m}(1 - \cos\theta)$$

$$\Delta G \text{ (energy)} = \sigma(surface\ energy) * \Delta A_{b=10\ \mu m}$$

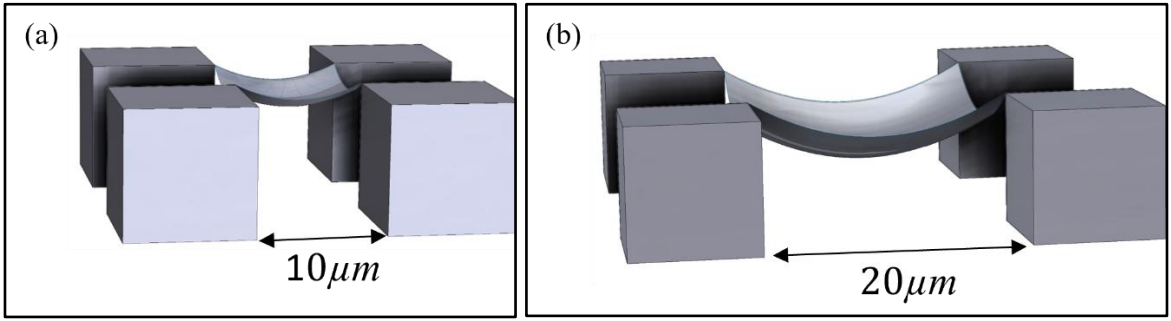

Figure S6. Schematics of the post with a thin layer (environment liquid replacing impregnated liquid) on top (a) 10 μm, (b) 20 μm post spacing

Similarly for 20-micron post spacings sample the area and the associated energy can be expressed as

$$A_{b=20\ \mu m} = 2\pi r^2_{b=20\ \mu m}(1 - cos\theta)$$

$$\Delta G \text{ (energy)} = \sigma(surface\ energy) * \Delta A_{b=20\ \mu m}$$

In comparison with the 10 micron post spacing sample to the 20 micron post spacing sample, the estimated $r$ is less qualitatively. The area and the energy barrier expression are directly proportional. As a result, compared to a 20 micron post spacing sample, the computed energy for a 10 micron post spacing sample is lower.

$$r_{b=10\ \mu m} < r_{b=20\ \mu m}$$

$$\Delta G_{b=10} < \Delta G_{b=20}$$

**Supporting Image S7. Distance vs time plot for other post spacing impregnated with high and low viscous oils.**

The high-viscosity Krytox 1506 and low-viscosity GPL 102 oils were impregnated into the textured surfaces with various micron post spacings, say 15 and 20, followed by kept inside the cyclopentane. These distance vs. time plots give an overview of the instability between two immiscible fluids. The plot below (figure S7.1) corresponds to a 15-micron post-spacing sample. It illustrates how instability grows at subsequent time intervals. The experimental data points were fitted for several capillary imbibition scaling regimes. This represents the interaction of viscous and capillary force, which mostly dictates the dynamics of this process. Similarly, the plot below (figure S7.2) corresponds to a 20-micron post-spacing sample.

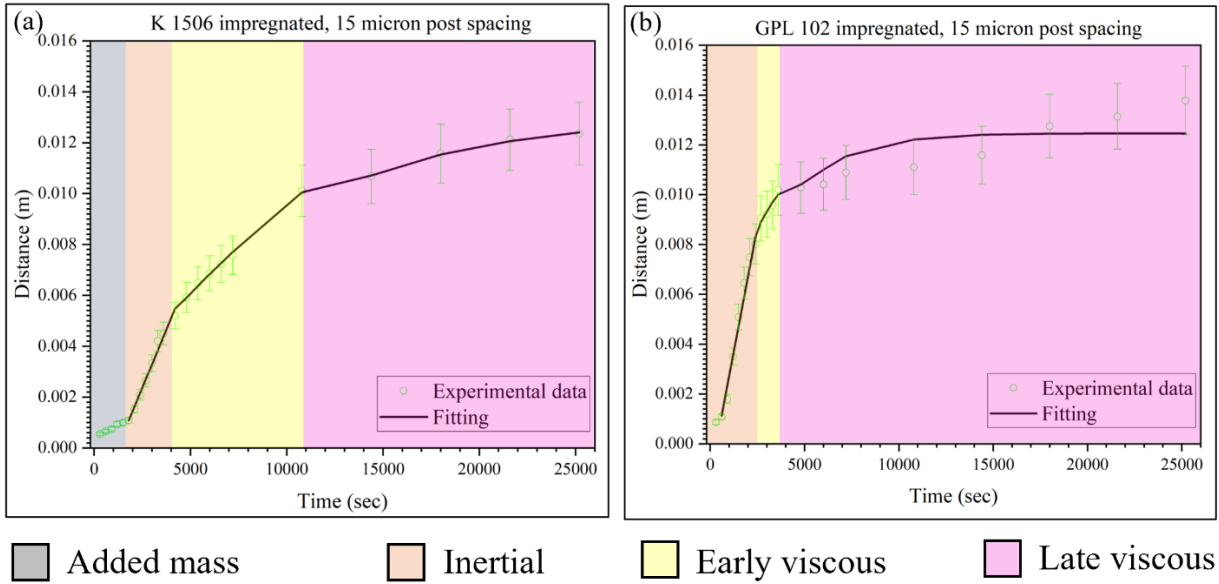

Figure S7.1 The various scaling regimes of capillary imbibition of distance vs time plot for (a) krytox 1506, (b) GPL 102 impregnated 15-micron sample under cyclopentane environment.

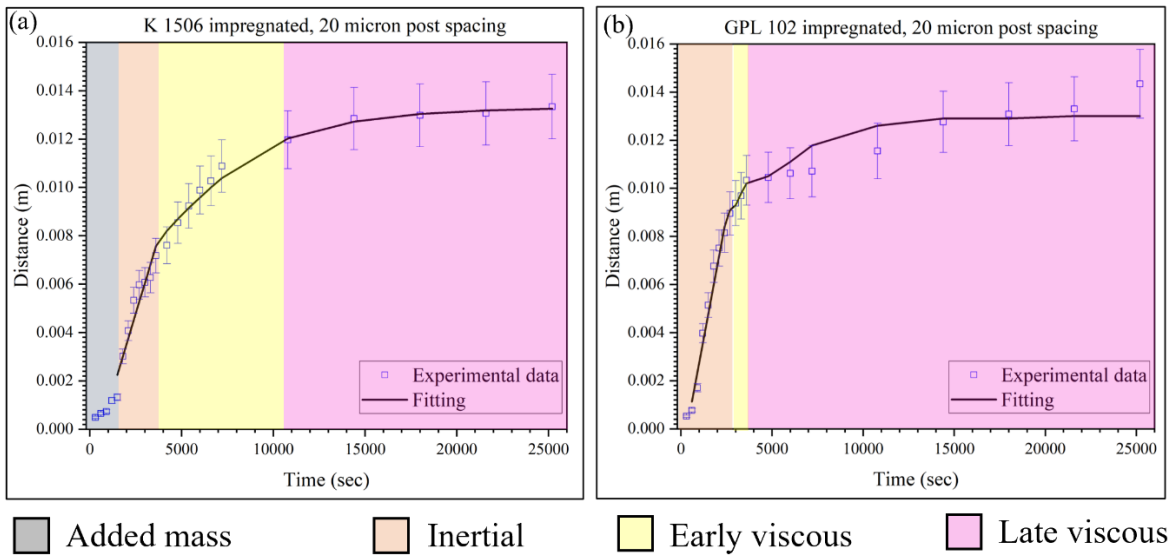

Figure S7.2. The various scaling regimes of capillary imbibition of distance vs time plot for (a) krytox 1506, (b) GPL 102 impregnated 20-micron sample under cyclopentane environment.